\documentclass[9pt,twocolumn]{extarticle}
\pdfoutput=1
\usepackage{soul}
\usepackage{titlesec} 

\usepackage[superscript, nomove]{cite}

\usepackage{float}
\usepackage{caption}
\usepackage{lipsum,ulem}
\usepackage[verbose]{placeins}

\usepackage{dsfont}
\usepackage{graphicx}
\usepackage{subcaption}
\usepackage[margin=0.9in]{geometry}
\usepackage[usenames,dvipsnames]{xcolor}
\usepackage[colorlinks,linkcolor=Blue,urlcolor=Blue,citecolor=Blue]{hyperref}
\usepackage{amsmath,amssymb}
\usepackage[squaren]{SIunits}

\usepackage{mathpazo}
\usepackage{courier}
\normalfont
\usepackage[T1]{fontenc}
\usepackage{caption}
\renewcommand{\thefigure}{\arabic{figure}}
\renewcommand{\figurename}{Fig.}
\usepackage{upgreek}

\newcommand{\ad}[1]{\textsuperscript{#1}\kern-2pt}

\makeatletter
\def\blx@maxline{77}
\makeatother

\usepackage{capt-of}

\def\mytitle{Pomeranchuk-like electronic localization above 100 K in twisted MoS$_{2}$}
\title{\vspace{-1.0cm}\huge\textbf{\textrm{\mytitle}}}  
\author{Zhiren Xiong,$^{1,2*}$ Jianqi Huang,$^{3*}$ Ruyue Han,$^{3}$ Hanwen Wang,$^{3}$ Hui Ding,$^{3}$ Jianming Lu,$^{3}$ \\ Jianpeng Liu,$^{4,5}$ Zheng Vitto Han,$^{1,2,3\dagger}$ Xingdan Sun,$^{3\dagger}$ Siwen Zhao,$^{3\dagger}$ Baojuan Dong$^{1,2,3,6\dagger}$}
\date{} 
\begin{document}
	\twocolumn[{
		\maketitle 
		\vspace{-5mm}
		\begin{center}
			\begin{minipage}{1\textwidth}
				\begin{center}
					\textit{
						\\\textsuperscript{1} State Key Laboratory of Quantum Optics and Quantum Optics Devices, Institute of Optoelectronics, Shanxi University, Taiyuan 030006, China
						\\\textsuperscript{2} Collaborative Innovation Center of Extreme Optics, Shanxi University, Taiyuan 030006, China
						\\\textsuperscript{3}Liaoning Academy of Materials, Shenyang 110167, China
						\\\textsuperscript{4}School of Physical Science and Technology, ShanghaiTech University, Shanghai 200031, China
						\\\textsuperscript{5}ShanghaiTech Laboratory for Topological Physics, ShanghaiTech University, Shanghai 200031, China
						\\\textsuperscript{6}Hefei National Laboratory, Hefei 230088, P. R. China
						\vspace{5mm}
						\\{$\dagger$} Corresponding to: vitto.han@gmail.com,  siwenzhao0126@gmail.com, xdsun@lam.ln.cn, dongbaojuan@sxu.edu.cn 
						\\{$*$} These authors contribute equally.
						\vspace{5mm}
					}
				\end{center}
			\end{minipage}
		\end{center}

\setlength\parindent{13pt}
\begin{quotation}
\noindent 
\section*{Abstract}
{\textbf{Twisted transition-metal dichalcogenides (TMDs) have manifested a rich variety of emerging physical phenomena, yet experimental studies have so far been largely limited in their valence bands (p-doped). Here, we show correlated electronic states in the conduction bands (n-doped) of near-AA-twisted bilayer MoS$_2$ with twist angles ranging from $\sim2.3^\circ$ to $\sim3.5^\circ$, down to the mK temperature regime. A strongly reconstructed correlated phase diagram as a function of twist-angle has been observed -- correlated gaps persist to temperatures approaching $160$ K at small twist angles,  but collapse to only $\sim20$ K at intermediate angles, where a richer landscape of interaction-driven states emerges. At the largest twist-angle $\sim 3.5\,^{\circ}$, correlated resistance at 1 electron per moir$\mathrm{\acute{e}}$ unit cell is enhanced upon heating, consistent with thermally assisted localization, or, a Pomeranchuk-like behaviour. Its magnetic-field response, however,  is highly anisotropic, which differs markedly from that of canonical isospin moir$\mathrm{\acute{e}}$ Pomeranchuk effect in graphene systems. Strikingly, such signature can persist even above 100\,K around a filling of 2 electrons per moir$\mathrm{\acute{e}}$, despite of its weak resistive nature. Our results establish the twisted MoS$_2$ as a platform for studying the complexity of charge localization, internal flavour degrees of freedom, and band topology in conduction bands of semiconducting moir$\mathrm{\acute{e}}$ systems.}}
\end{quotation}
}]

\newpage 
\clearpage

\section*{Introduction}

Moir\'e superlattices formed in van der Waals heterostructures modify microscopically the electronic landscape and promote Coulomb interaction within the resulting narrow bands. They have yielded, together with certain symmetry breaking, profound emergent physics including fractional Chern insulators and flat band superconductivity \cite{Cao2018Correlated,Cao2018Superconductivity,Tang2020,Regan2020,Wang2020TMD,Xu2020Fractional,Ghiotto2021Criticality,
Xu2022BilayerHubbard,Foutty2024Topology,Zhang2020FlatBands,Xia2025WSe2SC,Guo2025WSe2SC}. Among those moir\'e materials, twisted transition-metal dichalcogenides (TMDs) are particularly attractive because their semiconducting bands with spin-orbit coupling and locked spin-valley provide additional tuning knobs to the graphene-based moiré systems \cite{Xiao2012SpinValley,Naik2018Ultraflatbands,Zhang2020FlatBands,Devakul2021MagicTMD}. These internal degrees of freedom, together with strong layer hybridization and often electrically tunable moiré potentials, substantially enrich the accessible many-body phase space.

Despite rapid progress in the study of twisted TMDs, experimental investigations of twisted MoS$_{2}$ have remained rather limited, with most work focused on optical spectroscopy, scanning-probe measurements, or transport at elevated temperatures  \cite{Wang2020TMD,Ghiotto2021Criticality,Xu2022BilayerHubbard,Foutty2024Topology,Xia2025WSe2SC,Guo2025WSe2SC}. It is known that the intricate spin-valley structures, interlayer hybridization and moiré potential, together with predicted nontrivial topology \cite{Sharma2024, tMoS2_PRB_2020, PRResearch_Chern_2022}, make the MoS$_{2}$ conduction band of special interest, rather than a simple electron-hole counterpart of its valence band. In deed, recent studies have established moiré minibands and correlated insulating behaviour in electron-doped MoS$_{2}$ bilayers, highlighting the importance of layer hybridization and interaction-driven flavour symmetry breaking \cite{Wu2023MoS2,Zong2025MoSe2}. Yet systematic exploration of the correlated phase diagrams in near-AA twisted MoS$_{2}$ conduction band across different twist angles and down to the mK temperatutre regime have not been possible. A central obstacle is the difficulty of forming stable, ohmic metal contacts to MoS$_2$ at cryogenic temperatures. Nevertheless, our previous work overcame this limitation using a windowed-contact architecture, which enables reliable low-temperature transport in hexagonal boron nitride (h-BN) encapsulated MoS$_2$ devices \cite{ne_siwen}.

In this study, we investigate correlated electronic states in the conduction bands of near-AA-twisted bilayer MoS$_2$ with twist angles ranging from approximately $2.3^\circ$ to $3.5^\circ$. We find that the conduction-band correlated phase diagram undergoes a pronounced non-monotonic reconstruction with respect to the twist angle. At relatively small twist angles, correlated states at integer fillings persist to temperatures approaching $160$ K; at intermediate angles, their characteristic temperature collapsed to $\sim20$ K, whilst the system manifests a richer landscape of correlated states at both integer and fractional moir\'e fillings. At the largest twist angle $3.5\,^{\circ}$, the correlated resistance,
at the filling of one electron per moir$\mathrm{\acute{e}}$ unit cell, turns ultimately into an unconventional state in which thermal fluctuations reinforce localization rather than destroy it. Such a pronounced Pomeranchuk-like character is nearly insensitive to an in-plane magnetic field but is strongly enhanced by a perpendicular field, indicating an internal structure with spatial anisotropy, distinct from that of conventional nearly isotropic spin-/isospin-dominated moir$\mathrm{\acute{e}}$ Pomeranchuk systems observed in magic-angle twisted bilayer graphene. Even more strikingly, at the filling of two electrons per moir$\mathrm{\acute{e}}$ unit cell, a Pomeranchuk-like correlated signature persists above $100$ K. The effective mass extracted from low-temperature resistance measurements is as large as 4-8 times of bare electron's mass, indicating an emergent heavy-fermi-liquid behavior. Our results thus suggest that n-doped twist MoS$_2$ might be a unique platform that hosts unconventional tunability of moir$\mathrm{\acute{e}}$ band topology.

\section*{Results and Discussion}
\noindent\textbf{Fabrications and characterizations of t-MoS$_{2}$ devices.} 
\\
Monolayer crystalline MoS$_{2}$ flakes were prepared by exfoliating their bulk under ambient conditions. A dry-transfer method was then adopted to encapsulate twisted MoS$_{2}$ within two flakes of hexagonal boron nitride (h-BN) (more details can be found in Methods). Standard nano-fabrications were carried out to realize dual-gated devices  suitable for cryogenic transport in electron-doped twisted bilayer MoS$_2$. Opitcal micrograph of a representative device is shown in Fig. 1a. The active twisted MoS$_2$ region is contacted through a two-dimensional windowed-contact geometry with the semiconducting channel gated by top and bottom gates. As illustrated in Fig. 1b, the contacts are made by thermal-evaporated Bi/Au metallization through pre-patterned hollow windows in the top h-BN. This architecture yields reproducible ohmic contacts over a wide temperature range, from the millikelvin regime to room temperature, and enables systematic transport measurements in the conduction band of twisted MoS$_2$ (details including such as typical n-type field effect curves as well as $I$-$V$ characteristics in the studied devices can be seen in Supplementary Figure 1).

We first compare the zero-field transport at cryogenic temperatures of five representative near-AA-twisted bilayer MoS$_2$ devices with twist angles between $2.32^\circ$ and $3.56^\circ$ (Fig. 1c-g). The longitudinal resistance is plotted in the parameter space of displacement-field $D$ and carrier density $n$, with the density also expressed in terms of the moiré filling factor $\nu$ (defined as electrons per moiré unit cell area). Here, the effective displacement field $D_\mathrm{eff}=(C_\mathrm{tg}V_\mathrm{tg}-C_\mathrm{bg}V_\mathrm{bg})/2\epsilon_{0} - D_{0}$ can be applied between the top and bottom gates, and at the same time the total carrier $n_\mathrm{tot}=(C_\mathrm{tg}V_\mathrm{tg}+C_\mathrm{bg}V_\mathrm{bg})/e-n_{0}$ induced by the dielectrics from top and bottom gates can be tuned, as previously used in multiple-layered graphene devices \cite{FengWang_Nature_BLG,Maher_Science,Dong2026NatureSensors,sund2021}. Here, $C_\mathrm{tg}$ and $C_\mathrm{bg}$ are the top and bottom gate capacitances per area, respectively. And $V_\mathrm{tg}$ and $V_\mathrm{bg}$ are the top and bottom gate voltages, respectively. $n_{0}$ and $D_{0}$ are residual doping and residual displacement field, respectively. Notice that $D_{0}$ can not be easily determined definitively in the current system, we therefore include it in  $D_\mathrm{eff}$, written as $D-D_{0}$. Meanwhile, $n$ (also $\nu$) can be well defined using cross-checking of h-BN dielectric as well as Hall measurements and Landau fan diagrams at finite magnetic fields in each device (Supplementary Figures 2-6).

 \begin{figure*}[ht!]
 	\centering
 	\includegraphics[width=0.9\linewidth]{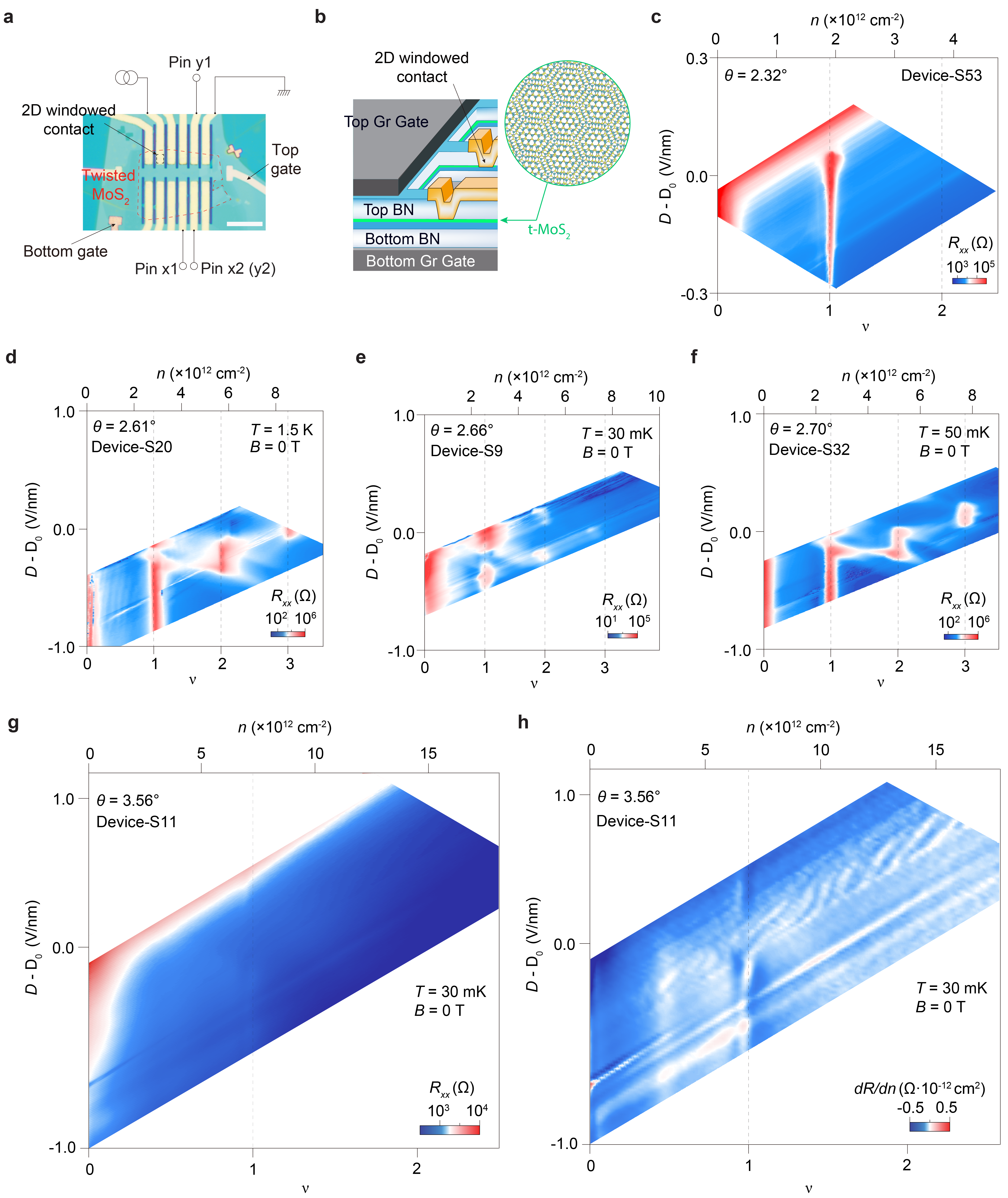}
     \caption{
    	\textbf{Device architecture and twist-angle evolution of correlated transport in	electron-doped near-AA-twisted bilayer MoS$_2$.} \textbf{a}, Optical micrograph of a representative twisted bilayer MoS$_2$ device fabricated with two-dimensional windowed contacts. \textbf{b}, Schematic of the dual-gated device structure. The windowed-contact geometry provides direct electrical access to the MoS$_2$ channel while preserving encapsulation, enabling stable ohmic transport from millikelvin temperatures to room temperature. \textbf{c--g}, Zero-field maps of the longitudinal resistance $R_{xx}$ as a function of displacement field $D$ and carrier density $n$, or equivalently the moiré filling factor $\nu$, for devices with twist angles of $2.32^\circ$ (Device S53), $2.61^\circ$ (Device S20), $2.70^\circ$ (Device S32), $2.66^\circ$ (Device S9) and $3.56^\circ$ (Device S11), respectively.
    	The measurements were performed at the temperatures indicated in each panel. The correlated resistance features evolve strongly with twist angle. The $2.44^\circ$ and $4.51^\circ$ devices predominantly exhibit a resistance
    	anomaly at $\nu=1$, whereas additional integer-filling features are resolved in the intermediate-angle devices.
    	\textbf{h}, Density derivative $dR_{xx}/dn$ of the $4.51^\circ$ data in \textbf{g}, which indicates the correlated states at $\nu$=1 of the lowest moiré miniband.
    }
	\label{fig:fig1}
 \end{figure*}

  \begin{figure*}[ht!]
  \centering
 	\includegraphics[width=0.95\linewidth]{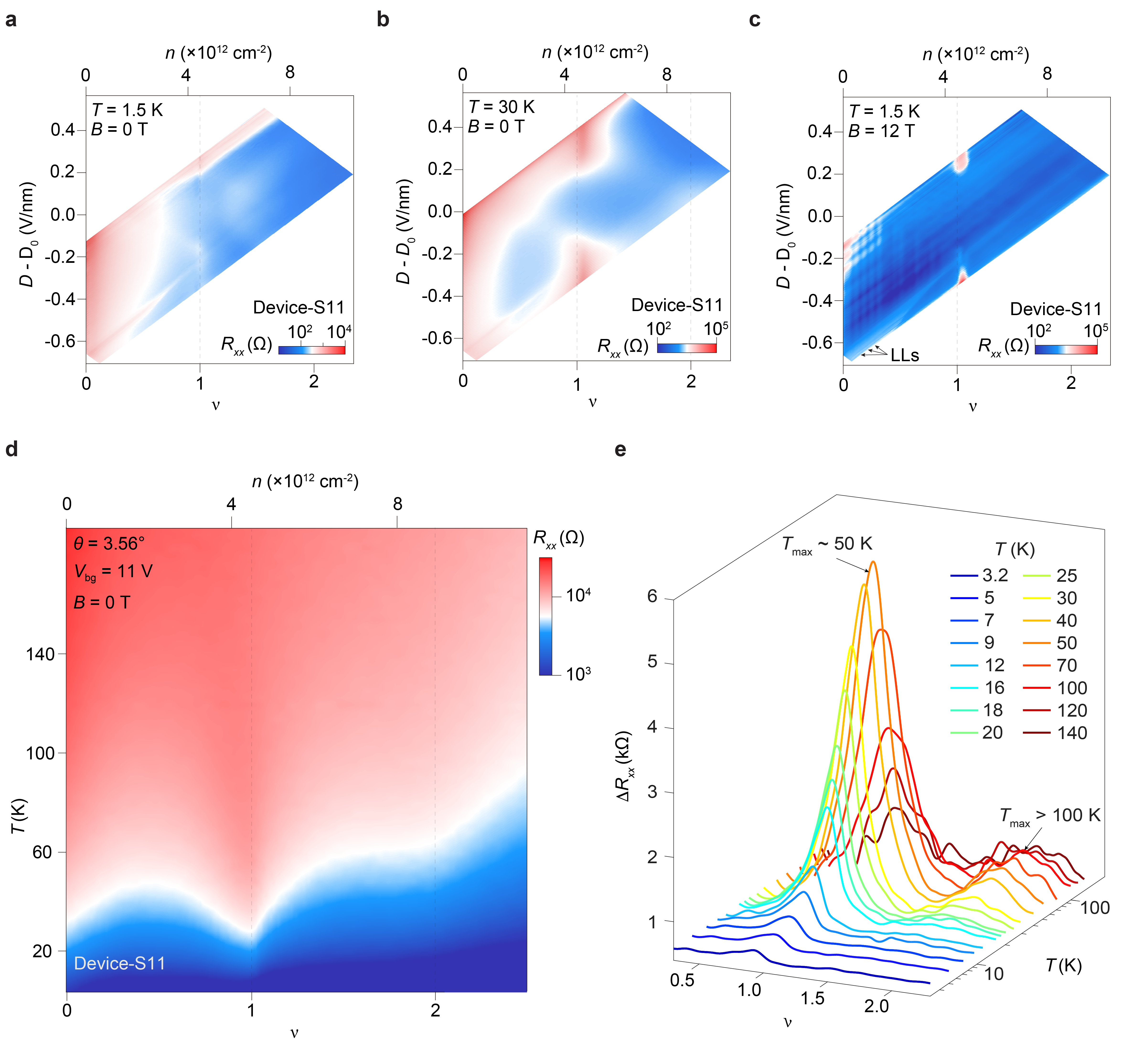}
 	\caption{\textbf{Thermally enhanced correlated resistance in the $3.56^\circ$ twisted bilayer MoS$_2$ device.} \textbf{a}, Zero-field $R_{xx}(D,\nu)$ map of Device S11 ($\theta=3.56^\circ$) measured at $T=1.5$ K. \textbf{b}, The corresponding map measured at $T=30$ K and $B=0$ T.  \textbf{c}, $R_{xx}(D,\nu)$ map measured at $T=1.5$ mK and $B=12$ T. Landau-level-like features are resolved, as indicated by the arrows. \textbf{d}, Temperature-filling-factor map of $R_{xx}$ at zero magnetic field. \textbf{e}, Line cuts at selected temperatures extracted from \textbf{d}, with a polynomial background subtracted from each curve.
 	}
 	\label{fig:fig2}
 \end{figure*}

   \begin{figure*}[ht!]
  \centering
 	\includegraphics[width=0.9\linewidth]{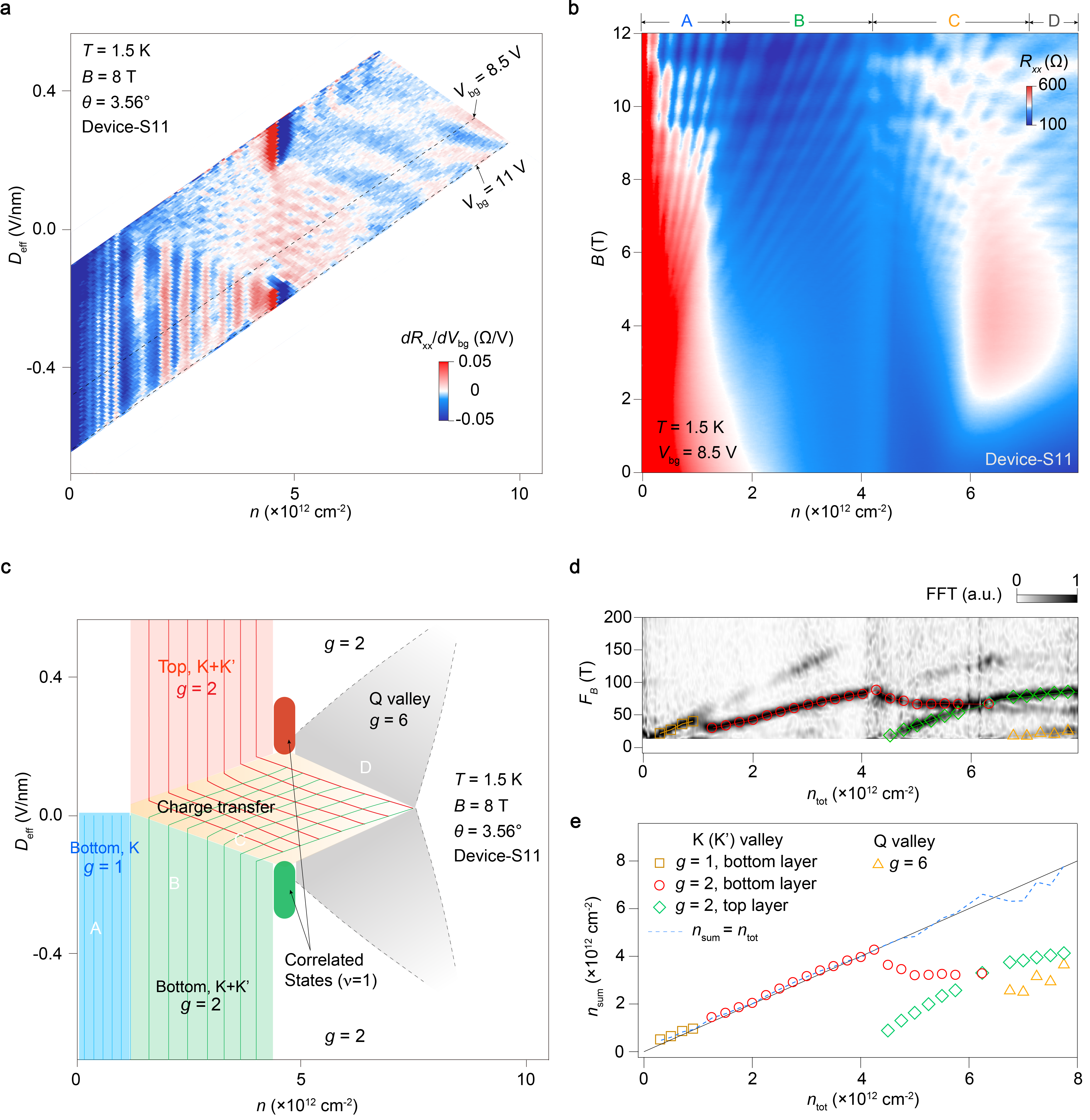}
 	\caption{\textbf{Layer and valley occupations in twisted MoS$_2$.} \textbf{a,} $\partial R_{\mathrm{xx}}/\partial V_{\mathrm{bg}}$ map of Device-S11 ($\theta=3.56^\circ$) as a function of carrier density $n$ and effective displacement field $D_{\mathrm{eff}}$ at $T=1.5$ K and $B=8$ T. \textbf{b,} Landau fan measured along the fixed-$V_{\mathrm{bg}}=8.5$ V trajectory indicated by the black dashed line in \textbf{a}. Labels A-D denote the occupation regimes identified at $B=12$ T. \textbf{c,} Schematic layer and valley occupation map inferred from the Landau-level features and quantum-oscillation analysis. Coloured regions indicate the assigned layer and valley degrees of freedom and their effective degeneracies $g$. The central region exhibits interlayer charge transfer; solid coloured shadows mark the correlated-state regions. \textbf{d,} Fourier spectra of the quantum oscillations versus inverse magnetic field. Symbols track the frequency branches assigned to bottom-layer $K/K'$ states, top-layer $K/K'$ states and Q-derived states. \textbf{e,} Carrier densities calculated from the marked branches using $n_i=g_i eF_i/h$. The blue dashed curve shows their sum, $n_{\mathrm{sum}}$; the black solid line indicates $n_{\mathrm{sum}}=n_{\mathrm{tot}}$.
 	}
 	\label{fig:fig3}
 \end{figure*}

\begin{figure*}[ht!]
  \centering
 	\includegraphics[width=0.9\linewidth]{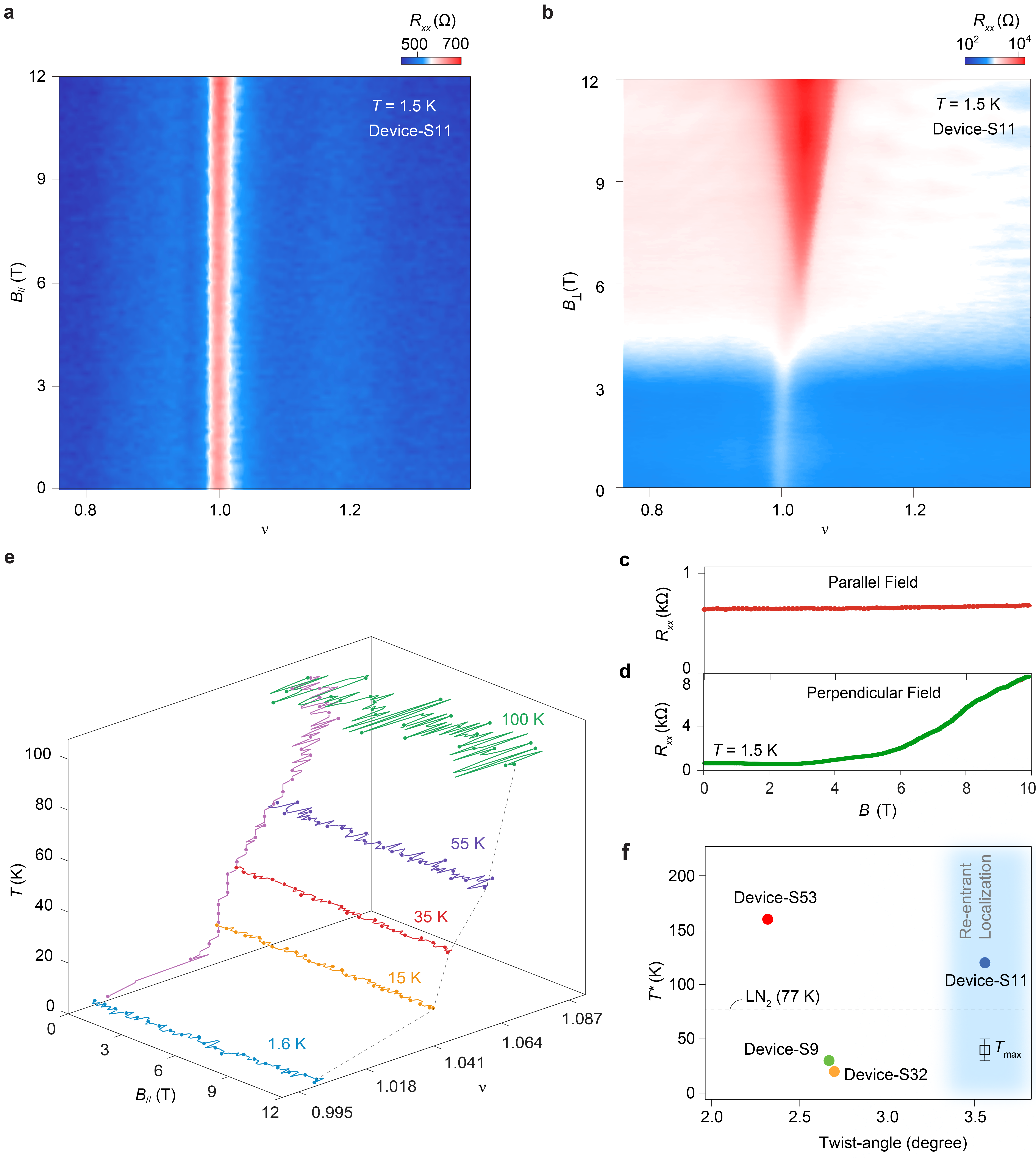}
 	\caption{\textbf{Magnetic-field anisotropy of the Pomeranchuk-like localized state.}
    \textbf{a,} Longitudinal resistance $R_\mathrm{xx}$ of Device-S11 at $T=1.5$ K as a function of filling factor $\nu$ and in-plane magnetic field $B_{\parallel}$, showing that the correlated resistance feature near $\nu=1$ is nearly insensitive to $B_{\parallel}$ up to 12 T. \textbf{b,} Corresponding $R_\mathrm{xx}(\nu,B_{\perp})$ map measured under a perpendicular magnetic field, where the $\nu=1$ resistance peak is strongly enhanced with increasing $B_{\perp}$. \textbf{c,d,} Evolution of $R_\mathrm{xx}$ at $\nu=1$ as a function of $B_{\parallel}$ and $B_{\perp}$, respectively, highlighting the pronounced magnetic-field anisotropy of the correlated state. \textbf{e,} Temperature-dependent evolution of the $\nu=1$ resistance feature in the three-dimensional parameter space of filling factor, temperature and in-plane magnetic field. The correlated localization remains largely insensitive to $B_{\parallel}$ over the temperature range in which the resistance anomaly is thermally enhanced. \textbf{f,} Characteristic temperature scale $T^{*}$ extracted from devices with different twist angles. The correlated states persist to $\sim160$ K at the smallest twist angle, are strongly suppressed to $\sim20$-30 K at intermediate angles, and re-enter a high-temperature localized regime at the largest twist angle. The dashed line marks the liquid-nitrogen temperature of 77 K, and the shaded region denotes the regime of re-entrant Pomeranchuk-like localization.}
 	\label{fig:fig4}
 \end{figure*}

Clearly, a pronounced non-monotonic angle-dependence is seen in the five representative devices shown in Fig. 1c-g. Although they all manifest density-dependent resistance anomalies - Coulomb interaction $U(\theta)$-driven gaps in a relatively flat band width $W(\theta)$. However, their number, position and displacement-field dependence vary substantially with respect to the twist angle $\theta$. Among them, the $2.32^\circ$-twisted device shows a dominant feature near $\nu=1$, with no resistance peaks at higher integer fillings within the measured density range. The correlated states evolves into a richer filling numbers of $\nu=1, ~2, ~3$ from $2.32^\circ$- to intermediate twist angles ($2.6 - 2.7^\circ$). At the largest twist-angle, i.e., $3.56^\circ$, notable moir\'e filling feature is seen only for $\nu=1$ again, despite its substantially different overall resistance landscape. Notably, the $2.32^\circ$-twisted device exhibits a correlated thermal scale $T^{*}$ up to 160\,K (Extended Data Figure 1), at which temperature the correlated $\nu=1$ resistive peak feature fades out, benchmarking a very strong localization against thermal activation in AA-twisted MoS$_{2}$. It is noticed that, in AB-twisted MoS$_{2}$, even higher $T*$ was reported therein \cite{Wu2023MoS2}. The weak density-dependent structures in the $3.56^\circ$ device are more clearly resolved by taking the density derivative of the resistance, highlighting the  $\nu=1$ anomaly in Fig. 1h. Notice that all the twist angles of the above devices are calibrated using cross-check of the corresponding carrier densities for each correlated peak (Supplementary Figures 2-6). The conduction-band moir\'e spectrum and its interaction-driven states undergo a substantial reconstruction across different AA-stacked twist angles. Evaluation of correlation strengths of several typical t-MoS$_{2}$ can be seen in Extended Data Figure 2a.

\vspace{5mm}
\noindent\textbf{Mapping the correlated states in near-AA-twisted MoS$_2$.} 
\\
We next examine in more details the correlated states in a near-$2.61^\circ$ twisted bilayer MoS$_2$ device (Device-S20). Extended Data Figure 3a shows the $R_{xx}(D,n)$ map of Device-S20 ($\theta=2.61^\circ$) at $T=1.5$ K and under a perpendicular magnetic field of $B=12$ T. The resistance map contains pronounced resistive peaks (red color) near each integer moiré fillings around $\nu=1, ~2, ~3$. At fixed displacement field, the magnetic-field dependence of the same device is displayed in Extended Data Figure 3b. Several high-resistance peak features evolve continuously in the manner of a fan-like structure with magnetic field, as expected for single-particle Landau levels. Meanwhile, additional magnetic-field independent resistive peaks are also seen for relatively lower fields in the range of 0 - 4 T in between filling 0 and 1. 

As shown in Extended Data Figure 3c, resistance line cuts obtained at $D_{\rm eff}=0.63$ V$\cdot$nm$^{-1}$ exhibit anomalies at the fractional fillings $\nu=3/11$, $3/8$, $7/15$, $4/7$, $2/3$ and $4/5$. These features occur at well-defined fractions of the moiré density and persist over a finite range of low magnetic fields. Their systematic evolution with $B$ and their commensurability with the moiré filling suggest that they may arise from interaction-driven charge ordering rather than from a simple single-particle density-of-states effect. We therefore refer to these states as fractional-filling charge-ordered states, while leaving their microscopic ordering pattern for further discussions. Similar features are indeed seen in other twisted TMDs systems, such as MoSe$_{2}$ moiré conduction band \cite{Zong2025MoSe2}.

Extended Data Figure 3d further shows resistance line cuts at $\nu=1$ and $\nu=2$ for a second near-$2.6^\circ$ Device-S32 (Device-S20 was damaged by discharge during measurement, and hence no further test available), measured at zero magnetic field and $D_{\rm eff}=-0.09$ V$\cdot$nm$^{-1}$ (see more temperature-dependent information in Supplementary Figure 7). Both integer-filling resistance peaks weaken continuously upon warming. They remain clearly resolved at low temperatures but are substantially broadened and suppressed at $T^{*}$ of about 25-30 K. 

The moiré correlations in samples with slight difference of twist-angle may yield distinct details. In the $2.66^\circ$ device (Device-S9), at $T=30$ mK and $B=10$ T, the density derivative $dR_{xx}/dn$ reveals a series of fine structures extending across the displacement-field--density plane (Extended Data Figure 4a, and Supplementary Figure 8). In particular, a set of evenly-spaced Landau-level-like features, not solely tuned by $n$ but also by $D$, appears in the region between $D$= -0.4 and 0 V/nm. Their strong dependence on displacement field identifies this region as an interlayer charge-transfer regime, in which the layer polarization of the electronic states evolves with carrier density and electrostatic bias. Further, the degeneracy between adjacent LLs are seen to be doubled for those LLs at moire fillings above $\nu$ =2, indicating the variation of valley flavor.

The magnetic-field evolution is shown more directly in Extended Data Figure 4b, where $R_{xx}$ is plotted as a function of $\nu$ and $B$ at $D_{\rm eff}=-0.3$ V$\cdot$nm$^{-1}$. And the thermal evolution of the Device-S9 is shown in Extended Data Figure 4c. As shown in Extended Data Figure 4d-e, the states at $\nu=1$, $2$ and $3$ display related but non-identical temperature dependencies. We notice that the zero magnetic field behaviors (Fig. 1d-f) are rather similar in Device-S20, S9, and S32, since the variation of twist angles between them are relatively small, the details of their transport signatures at finite magnetic fields are yet drastically distinct. It indicates a very sensitive response in terms of twist angle, and/or sample quality.

\vspace{5mm}
\noindent\textbf{Pomeranchuk-like thermal enhancement of localization in a $3.56^\circ$ moiré band.} 
\\
Centrual results of this study is the anomalous, thermal assisted electron localizations in $3.56^\circ$ twisted bilayer MoS$_2$ device (Device-S11), which exhibits a drastically different thermal response from that of those smaller-angle samples. Figure 2a shows the zero-field $R_{xx}(D,\nu)$ map measured at $T$ = 1.5 K for pin 4-5 (see more in Supplementary Figure 9, and data taken at mK for another pair of pin15-16 are seen in Fig. 1g). At this temperature, the resistance landscape is relatively smooth over the electron-doped region, with only a pair of faint peak near $\nu=1$ separated for positive and negative $D$. Upon increasing the temperature to $30$ K, however, a much more pronounced resistance feature develops around the same filling (Fig. 2b). In stark contrast to the conventional expectation of Mott-like correlated insulating state, the $\nu=1$ resistance peak here in Device-S11 gets enhanced and evolves to an insulating-like behavior only at elevated temperatures, indicating a heat-induced charge localization at such commensurate doping density. Shown in Fig. 2c, at $T$ = 1.5 K and $B$ = 12 T, the same correlated resistance peaks are comparatively stronger than their zero-field state. Additional Landau-level-like features also become visible. The temperature evolution of resistance between pin 1-2 at zero magnetic field is further shown in Fig. 2d. The resistance near $\nu$ = 1 increases upon warming over an intermediate temperature range, reaches a maximum $T_\mathrm{max}$ at around 50 K, and is subsequently suppressed at higher temperature. Line cuts extracted at representative temperatures after background subtracted (see Supplementary Figure 10 for details) are shown in Fig. 2e, exhibiting the non-monotonic evolution of the $\nu=1$ resistance peak. Strikingly, much weaker resistive peaks are found at $\nu=2$, but $T_\mathrm{max}$ reaches a magnitude above 100 K. 

Similar anomalous thermal-assisted localization effects in moiré superlattices were reported in magic-angle twisted bilayer graphene (TBG) \cite{Pomeranchuk_2021a,Pomeranchuk2021_MATBG,liuIsospin2022e}, as well as WSe$_{2}$/MoTe$_{2}$ heterostructures \cite{Pomeranchuk_prx}, often referred to as the Pomeranchuk-effect, which is an electronic analog of the higher-temperature solidification of $^{3}$He \cite{Richardson1997Pomeranchuk}. It is known that a correlated state hosting fluctuating local isospin degrees of freedom may carry a larger entropy than the competing itinerant fermi liquid state \cite{Pomeranchuk_2021a,Pomeranchuk2021_MATBG,PhysRevB.70.155114}. In this regime, increasing temperature can therefore enhance localization over a finite temperature range, producing a Pomeranchuk-like response before thermal fluctuations eventually destroy the correlated state. In magic-angle TBG, the low-energy electronic states are modeled by an emergent local degrees of freedom in the $AA$ region, coupled with the itinerant electrons in $AB$ region\cite{PhysRevLett.129.047601,PhysRevB.106.245129}. The latter was attributed to a Kondo screening of the emergent local moments by itinerant electrons, which produces a low-temperature heavy Fermi liquid metallic regime \cite{Zhou2024Kondo,Chou2023Kondo}. Upon heating, the suppression of Kondo correlations restores local-moment entropy, providing a microscopic interpretation of
the Pomeranchuk-like response in moir\'e systems \cite{Zhou2024Kondo}. We also fit the low-temperature resistance as $R_\mathrm{xx}(T)=R_\mathrm{xx}(0)+A T^2$, and an effective mass as large as $m^*\sim 4-8\,m_\mathrm{e}$ can be extracted from the coefficient $A$ (Supplementary Figure 11), significantly larger than that reported in magic-angle twisted trilayer graphene\cite{zzks-vkl2}. This suggests the emergence of a heavy fermi liquid in the $3.5^{\circ}$ device, which can be potentially described by a Kondo-lattice-like model emerging on the moir\'e length scale, reminiscent to the scenario of magic-angle graphene. The emergence of local quantum degrees of freedom may be attributed to the fact that the Ising spin-orbit coupling (SOC) in the $K$-valley conduction band of twisted MoS$_2$ is only a few meV (see Extended Data Figure 2b), two orders of magnitude smaller than that of the valence bands. This leaves the physical spin of each valley as a quantum degree of freedom rather than being frozen. In addition to the spin-valley degeneracy, electrons are localized at XM and MX sites, forming a layer-pseudospin degree of freedom (see Extended Data Figure 2c--e). These multiple valley, spin and layer quantum degrees of freedom can all contribute to the entropy gain, favoring the emergence of local moments upon heating.

\vspace{5mm}
\noindent\textbf{Determining the flavors of degeneracy in $3.56^\circ$ t-MoS$_{2}$.} 
\\ Before exploring the magnetic-field dependence, we first analyze the flavor degeneracy of the system at 3.56$^{\circ}$, which may provide further information about the origin of the Pomeranchuk-like effect. It is seen from Fig. 2c that the $D$-$\nu$ map of Device-S11 at $T$= 1.5 K and $B$= 12 T shows Landau level features (indicated by solid arrows in Fig. 2c). It provides the possibility to determine in detail the degeneracies of spin, valley and layer in the current system. Figure 3a shows the  $D$-$\nu$ map of Device-S11 at $T$= 1.5 K and $B$= 8 T, with the $R_\mathrm{xx}$ data plotted in the form of derivative of $V_\mathrm{bg}$ for visual clarity. The stripe-like LLs features in the lower-left part of the map evolve into a checkerboard-like crossing network in the central region. This pattern is consistent with crossings between LLs associated with the two constituent layers, accompanied by interlayer charge redistribution, as also observed in large-angle twisted graphene \cite{Dong2026NatureSensors}. The crossing network indicates that the relevant states retain distinct layer character, consistent with the fact that the interlayer coupling is weak compared with the kinetic energy at relatively large twist angle. In this central region, labelled as C in Fig. 3b-c, the correlated-state signatures are no longer resolved. Figure 3b shows $R_{\mathrm{xx}}$ as a function of carrier density and perpendicular magnetic field along the fixed $V_{\mathrm{bg}}=8.5$ V trajectory marked by the black dashed line in Fig. 3a. Several LL sequences are resolved, including a range in which distinct sequences intersect. Their evolution, together with the quantum-oscillation analysis below, identifies four occupation regimes, labelled A-D along the $B=12$ T edge of Fig. 3b and summarized schematically in Fig. 3c.

To determine the effective degeneracies, we perform a fast Fourier transform of the oscillations as a function of inverse magnetic field (Fig. 3d). Figure 3e compares the densities associated with the individual branches and their sum, $n_{\mathrm{sum}}=\sum_i n_i$, with the gate-induced total density $n_{\mathrm{tot}}$. The approximate agreement between $n_{\mathrm{sum}}$ and $n_{\mathrm{tot}}$ supports the assignments in Fig. 3c. Region A is characterized by a single resolved flavour with $g=1$, assigned to a bottom-layer $K$-valley state. In region B, the dominant branch remains associated with the bottom layer but has $g=2$, consistent with contributions from both $K$ and $K'$ valleys. In region C, a second branch, assigned to top-layer valley-degenerated $K/K'$ states with $g=2$, emerges as the bottom-layer frequency decreases. At higher density, an additional low-frequency branch is accounted for by $g=6$ and assigned to Q-derived states, identifying the onset of $Q$-valley fillings in region D {\cite{masseroni2025gate,wu2016even}}. The proximity to boundaries separating different layer and valley occupations further indicates that the correlated states at $\nu$=1 at finite $D_\mathrm{eff}$ (indicated by red and blue solid areas in Fig. 3c) develop in a complex environment. In addition to the spin degrees of freedom within each valley, states with such multiple degenerate $K/K'$ and/or Q valleys may provide an entropy advantage to the correlated state over the competing itinerant state, thus favoring Pomeranchuk-like localization. This is plausible given that the Pomeranchuk-like localization, as shown by the $T$-$\nu$ mapping in Fig. 2d, emerges in regimes where the $K$ and presumably $Q$-valley intersect (see dashed line in Fig. 3a, $V_\mathrm{bg}$= 11 V).

\vspace{5mm}
\noindent\textbf{Magnetic anisotropy of the Pomeranchuk-like correlated states.} 
\\ 
To further reveal the possible spin or valley induced entropy in the system, we carried out in-plane and out-of-plane magnetic field measurements for $\nu=1$ ($\nu=2$ is too weak to be tested in a reasonable temperature range, since it is featureless below 50\,K). As shown in Fig. 4a, the correlated resistance peak at $\nu=1$ is remarkably insensitive to an in-plane magnetic field $B_{\parallel}$ up to 12 T. The corresponding line cut (in Fig. 4c) shows only a negligible variation of $R_{xx}$ over the entire field range. This behaviour contrasts sharply with the magnetic-field response reported in magic-angle twisted bilayer graphene, where the excess entropy associated with the Pomeranchuk-like state is substantially suppressed by an in-plane magnetic field, pointing to a pronounced isotropic spin or isospin origin \cite{Pomeranchuk_2021a,Pomeranchuk2021_MATBG,liuIsospin2022e}. In our twisted MoS$_2$ device, the absence of a sizeable $B_{\parallel}$ response therefore suggests that the thermally enhanced localization is unlikely to originate predominantly from conventional isotropic spin entropy.

By contrast, the same $\nu=1$ state responds strongly to a perpendicular magnetic field $B_{\perp}$. As shown in Fig. 4b, the resistance peak around $\nu=1$ becomes progressively enhanced with increasing $B_{\perp}$, while the corresponding line cut in Fig. 4d reveals a pronounced increase of $R_{xx}$ above several tesla. Such a marked field anisotropy indicates an unconventional origin of the local moment degrees of freedom: they either originate from orbital, valley, or other band-structure-related degrees of freedom, or from physical spins that have Ising anisotropy due to the non-negligible SOC ($\sim$ a few meV). This behaviour is also distinct from that reported in 3L-MoTe$_2$/WSe$_2$ moiré heterostructures, where an out-of-plane magnetic field drives a much stronger metal--insulator transition accompanied by quantum-critical scaling \cite{Pomeranchuk_prx}. The present results therefore point to a Pomeranchuk-like regime in twisted MoS$_2$ with a microscopic origin different from both the predominantly isotropic isospin-driven response in magic-angle graphene and the field-driven critical localization observed in MoTe$_2$/WSe$_2$. Figure 4e summarizes the evolution of the $\nu=1$ feature in the three-dimensional parameter space of filling factor, temperature and in-plane magnetic field, further demonstrating that the Pomeranchuk-like localization remains essentially unaffected by $B_{\parallel}$.

Finally, Fig. 4f summarizes the characteristic temperature scale $T^{*}$ extracted from devices with different twist angles. The correlated states survive to temperatures approaching 160\,K at the smallest twist angle, are strongly suppressed to only $\sim20$ - 30 K at intermediate twist-angles, and re-enter a high-temperature localized regime (in the meantime, a Pomeranchuk-like localization regime) at the largest twist angle. This strongly non-monotonic evolution highlights that twist angle reconstructs not only the strength of electronic correlations, but also the nature of the localized states in the conduction-band moiré system.

\vspace{5mm}

To conclude, we systematically investigate electron transport in near-AA-twisted bilayer MoS$_2$ over a range of twist angles (from approximately $2.3^\circ$ to $3.56^\circ$), carrier densities, displacement fields, temperatures and magnetic fields. We found that twist angle drives a pronounced non-monotonic reconstruction of correlation physics in the conduction bands of near-AA-twisted MoS$_2$, evolving from robust high-temperature localization, through an intermediate regime with strongly suppressed thermal scales, to a re-entrant regime in which increasing temperature counterintuitively reinforces localization. This Pomeranchuk-like thermal localization exhibits a pronounced magnetic-field anisotropy, pointing to internal degrees of freedom distinct from those underlying the predominantly spin- or isospin-driven Pomeranchuk physics established in other moir$\mathrm{\acute{e}}$ systems. Remarkably, at a filling of two electrons per moir$\mathrm{\acute{e}}$ unit cell in Device-S11, the Pomeranchuk-like correlated signature persists above $100$ K despite its comparatively weak resistive amplitude. Our results establish electron-doped twisted MoS$_2$ as a distinct semiconducting moir$\mathrm{\acute{e}}$ platform in which twist angle controls not only the strength, but also the nature and thermal stability, of correlated electronic states.


\clearpage

 \clearpage

\section*{Methods}
\vspace{3mm}
\noindent\textbf{Fabrication.} 
Monolayer MoS$_2$ and few-layer hBN flakes were mechanically exfoliated from high-quality bulk crystals. MoS$_{2}$ monolayers were mechanically cut/scratched via an atomic force microscope (AFM) tip into two or more pieces, before being twist-stacked, in order to obtain as precise as possible the interlayer rotated crystallographic orientations. The heterostructures were assembled using a PC/PDMS-based dry-transfer technique. A top hBN flake containing pre-etched windows of approximately $1 \times 1~\mu\mathrm{m}^2$ was first picked up, followed by one MoS$_2$ piece. After rotating the substrate by the target angle $\theta$, the second piece was picked up to form AA-stacked twisted bilayer MoS$_2$. The stack was then released onto a $20$--$30$-nm-thick bottom hBN flake covering a local Au back gate. A Cr/Au ($5/20~\mathrm{nm}$) top gate was first fabricated by electron-beam lithography, thermal evaporation and lift-off. Bi/Au ($25/30~\mathrm{nm}$) contact electrodes were subsequently deposited by thermal evaporation through the pre-etched windows in the top hBN, providing direct surface contacts to MoS$_2$. Finally, the hBN-encapsulated heterostructure was patterned into a Hall-bar geometry by electron-beam lithography and reactive-ion etching, yielding a dual-gated twisted bilayer MoS$_2$ device.

\vspace{3mm}
\noindent\textbf{Electrical measurements.} 
Room-temperature electrical characterization was performed using a Cascade M150 probe station equipped with an Agilent B1500A semiconductor device parameter analyzer. Gate voltages were applied using a Keithley 2400 source meter. For low-temperature magnetotransport measurements, the devices were excited with an a.c. bias current, $I_{\mathrm{bias}}$, of $5$ or $10~\mathrm{nA}$. Low-frequency lock-in measurements in a four-probe configuration were performed at various temperatures and magnetic fields.

\vspace{5mm}
\noindent\textbf{Theoretical simulations.}
The moiré miniband structure and charge density of near-AA-twisted MoS$_2$ under five specific commensurate twist angle (3.48$^\circ$, 3.89$^\circ$, 4.41$^\circ$, 5.09$^\circ$ and 6.01$^\circ$) are calculated using first-principles calculations as implemented in Vienna Ab-Initio Simulation Package (VASP). Projected-augmented-wave (PAW) pseudopotential and the generalized gradient approximation (GGA) of Perdew-Burke-Ernzerhof (PBE) functional are used. The cutoff energy for the plane wave basis is 350~eV, and the convergence criteria for the self-consistent calculation is set to 10$^{-6}$~eV. The spin-orbit coupling and interlayer van der Waals corrections are considered in the performance. A vacuum layer of 20~Å alone the $z$ direction is used in these systems to avoid spurious interactions between periodic images. All the crystal structures including atomic positions have been fully relaxed using the conjugate gradient method, until none of the residual Hellmann-Feynman forces exceeds 10$^{-2}$~eV/Å. We use the formula $U \approx e^2/(4\pi\varepsilon_0\,\varepsilon_r\,a_M)$ to estimate the on-site Coulomb energy, where $a_M$ is the lattice constant after relaxation and $\varepsilon_r=10$ the effective dielectric environment. $U/W$ ratio is used to characterize the correlation strength of the lowest moiré conduction band with $W$ the corresponding calculated bandwidth, which shrinks from 120~meV to 40~meV as the twist angle decreases. The flavor assignment of the eight lowest conduction bands is obtained from the spin-polarization and valley-unfolding analysis of the DFT wavefunctions.

\section*{\label{sec:level1}Data Availability}

The data that support the findings of this study will be available at Zenodo repository with a public doi link.

\section*{\label{sec:level2}Code Availability}
The code that support the findings of this study are available upon reasonable request to the corresponding authors.

\section*{\label{sec:level3}Acknowledgements}
This work is supported by the National Key R$\&$D Program of China (Grant Nos. 2024YFA1410400, 2022YFA1203903) and the National Natural Science Foundation of China (NSFC) (Grant Nos. 12622430, 12574074, 92265203, 12104462, 11974357, U1932151, 12204287, 12204490, and 11974027,12374185). Z.V.H. acknowledges the support of the Fund for Shanxi “1331 Project” Key Subjects Construction. Z.V.H. and B.D acknowledge supports from the Quantum Science and Technology-National Science and Technology Major Project (Grant No. 2021ZD0302003). X.S acknowledges support from Liaoning Provincial Natural Science Fund with Grant 2025-MS-053 and Shenyang Special Program for Science and Technology Talents (U35 Outstanding Young Talent Program RC240631).

\section*{Author Contributions}
B.D., S.Z., Z.V.H., and X.S. conceived the experiment and supervised the overall project. Z.X., S.Z, B.D., and X.S. performed the device fabrications and low-frequency electrical measurements; R.H., H.W. and H.D. helped in sample fabrications. Z.V.H., S.Z, B.D., X.S, J.L., and J.H. analyzed the experimental data. The manuscript was written by Z.V.H. S.Z, B.D., X.S., and Z.X. with discussion and inputs from all authors.

\section*{Competing Interests}
The authors declare no competing interests.

\clearpage

\setcounter{figure}{0}
\renewcommand{\figurename}{Extended Data Fig.}
\renewcommand{\thefigure}{\arabic{figure}}

\begin{figure*}[p]
\section*{Extended Data Figures}
  \centering
 	\includegraphics[width=0.9\linewidth]{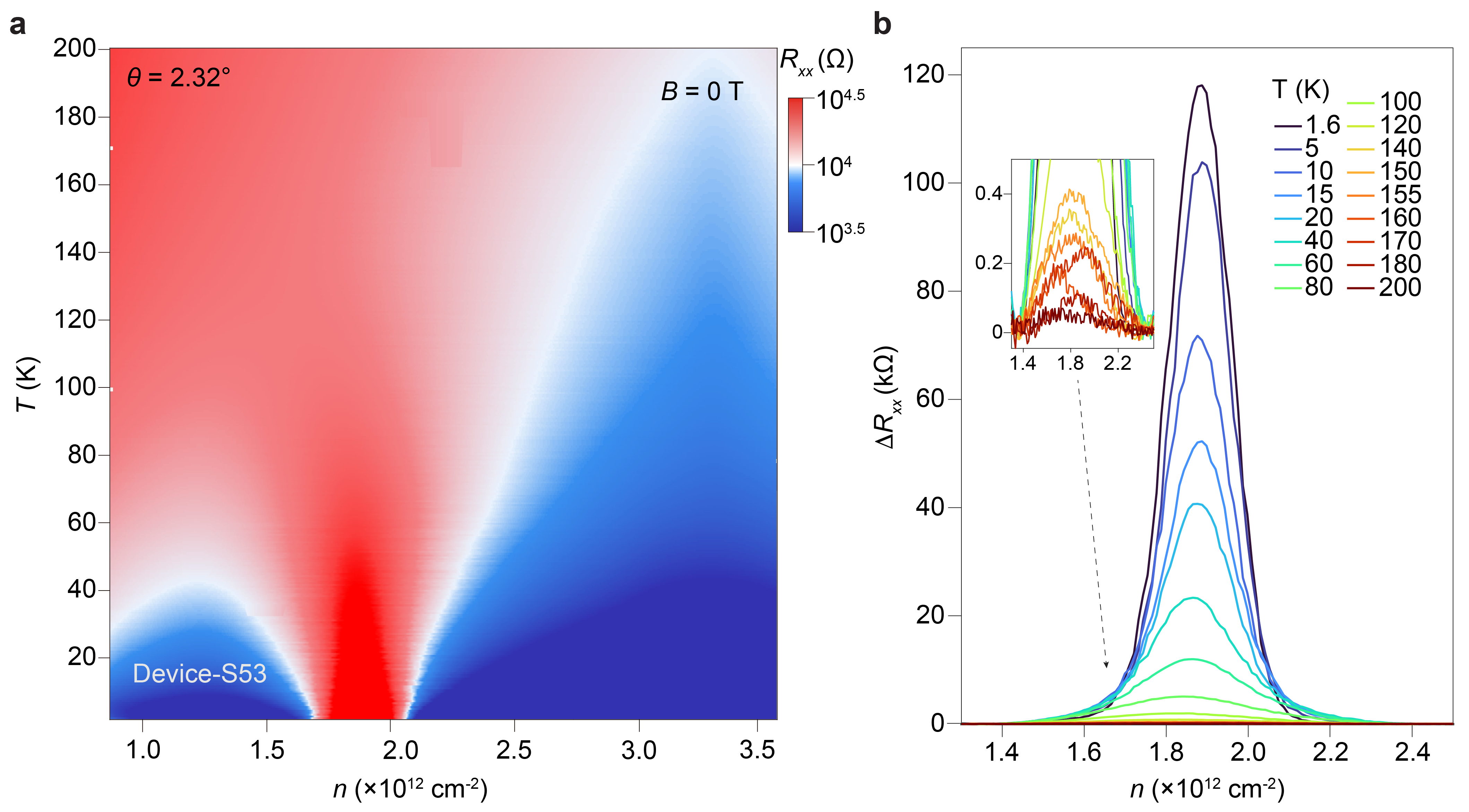}
 	\caption{
 		\textbf{Temperature dependence of correlated localization at $\nu$=1 in 2.32$^\circ$-twisted MoS$_{2}$.}
  \textbf{a,} Longitudinal resistance $R_{xx}$ of Device-S53 in the space of of density and temperature. The resistive peak at $n ~\sim~ 1.8 \times 10^{12} cm^{-2}$ corresponds to the filling of one electron per moire unit cell.  \textbf{b,} Line profiles of several characteristic temperatures in  \textbf{a}, with the polynomial background subtracted from each curve. The correlated state displays a pronounced resistive peak above 160 K.
 	}
 	\label{fig:extended-data-1}
 \end{figure*}

 \begin{figure*}[p]
 	\centering
 	\includegraphics[width=0.9\linewidth]{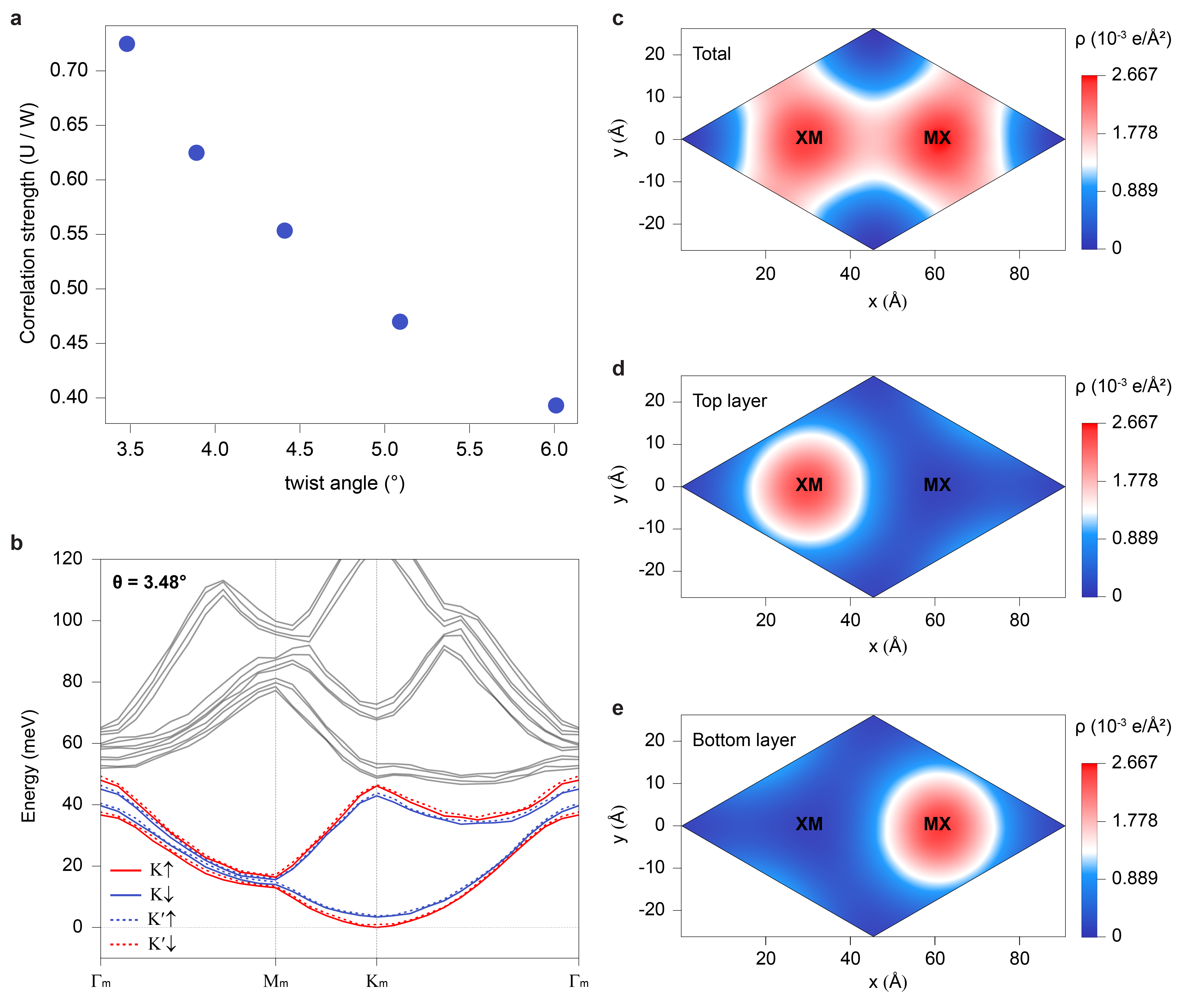}
    \caption{\textbf{Correlation strength and flavor structure of the moiré conduction bands in AA-stacked twisted bilayer MoS$_2$.} \textbf{a}, Correlation ratio $U/W$  of the lowest moiré conduction band with various twist angles, where $W$ is the bandwidth obtained from the DFT calculation, and $U$ is the on-site Coulomb energy estimated as the charging energy of a moiré unit cell. $U/W$ increases monotonically from 0.39 at 6.01$^\circ$ to 0.72 at 3.48$^\circ$, indicating a flattening band as the twist angles decrease. \textbf{b}, Flavor-resolved conduction-band structure at $\theta = 3.48^\circ$. The eight lowest conduction bands are classified into four valley--spin flavors $(\tau, s) = \mathrm{K}\uparrow$ (red solid), K$\downarrow$ (blue solid), K$^\prime\uparrow$ (blue dashed), and K$^\prime\downarrow$ (red dashed), each comprising two bands. \textbf{c}, Real-space charge-density distribution of the eight lowest conduction bands at the mBZ center $\Gamma_m$, integrated along the out-of-plane direction. The four flavor-resolved distributions are nearly identical to one another and to the total pattern shown, reflecting the approximate spin--valley degeneracy. \textbf{d}--\textbf{e}, The same density projected onto the two constituent layers, with the top (bottom) layer localized almost exclusively on the XM (MX) sites.
    }
    \label{fig:extended-data-2}
 \end{figure*}

 \begin{figure*}[p]
 	\centering
 	\includegraphics[width=0.9\linewidth]{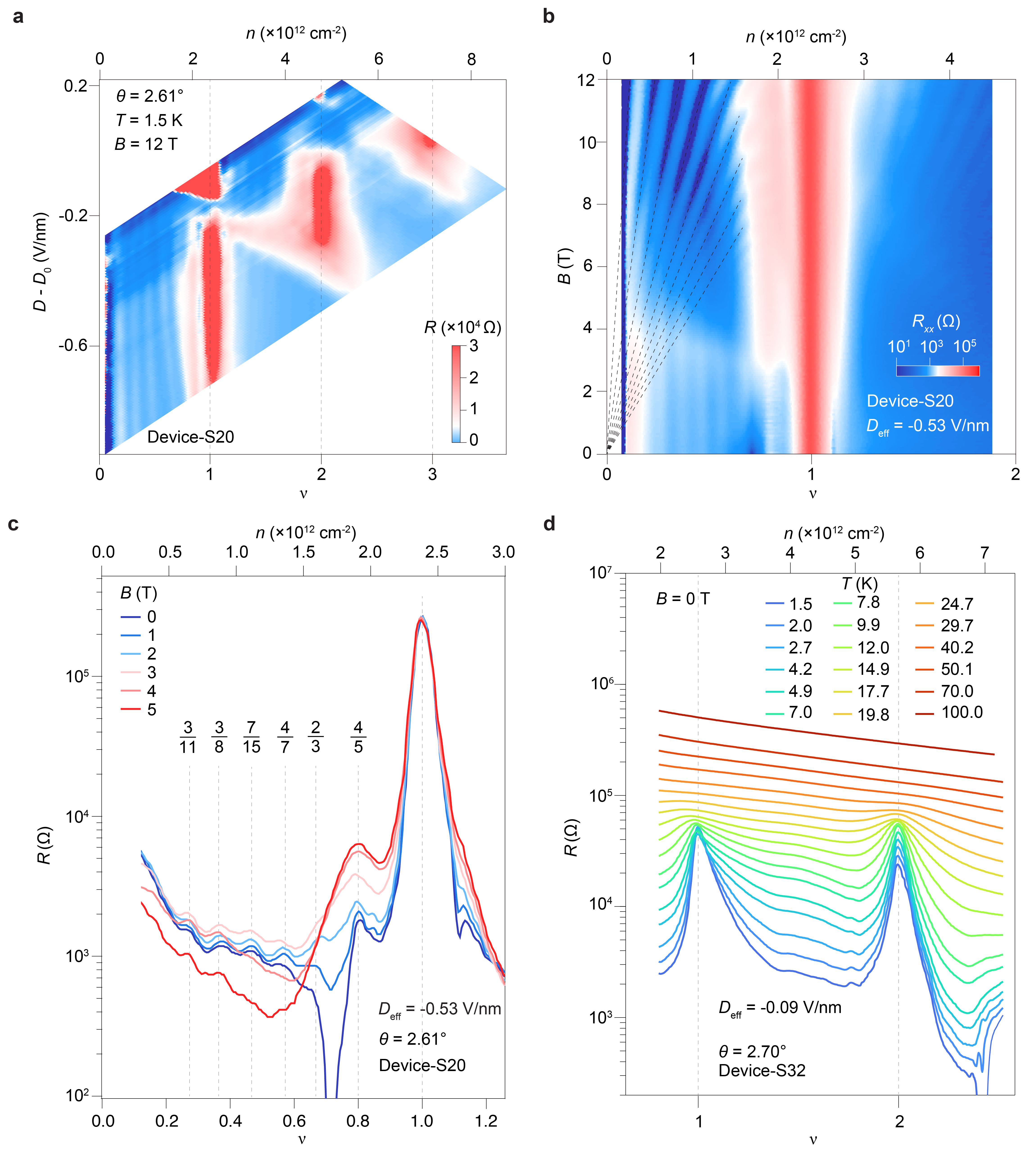}
   \caption{
   	\textbf{Magnetic-field evolution and thermal robustness of correlated states in $2.61^\circ$ twisted bilayer MoS$_2$.}
 		\textbf{a}, Longitudinal resistance $R_{xx}$ of Device S20 ($\theta=2.61^\circ$) mapped as a function of moiré filling factor $\nu$
 		and displacement field $D$ at $T=1.5$~K and $B=12$~T.
 		\textbf{b}, Landau-fan map of $R_{xx}$ for the same device as a function of
 		$\nu$ and perpendicular magnetic field $B$, measured at an effective
 		displacement field $D_{\rm eff}=0.63$ V$\cdot$nm$^{-1}$.
 		The dashed lines indicate the trajectories of the conventional Landau
 		levels.
 		\textbf{c}, Density line cuts extracted from the low-field data at
 		$D_{\rm eff}=0.63$ V$\cdot$nm$^{-1}$, showing a series of resistance anomalies at
 		fractional moiré fillings, including
 		$\nu=3/11$, $3/8$, $7/15$, $4/7$, $2/3$ and $4/5$.
 		These fractional-filling features are consistent with the formation of
 		interaction-driven charge-ordered states.
 		\textbf{d}, Temperature-dependent resistance line cuts at $\nu=1$ and $\nu=2$ in a near-$2.7^\circ$ device ($\theta=2.70^\circ$, Device S32) at
 		$D_{\rm eff}=-0.09$ V$\cdot$nm$^{-1}$ and $B=0$.
   }
 	\label{fig:extended-data-3}
 \end{figure*}

\begin{figure*}[p]
  \centering
  \includegraphics[width=0.9\linewidth]{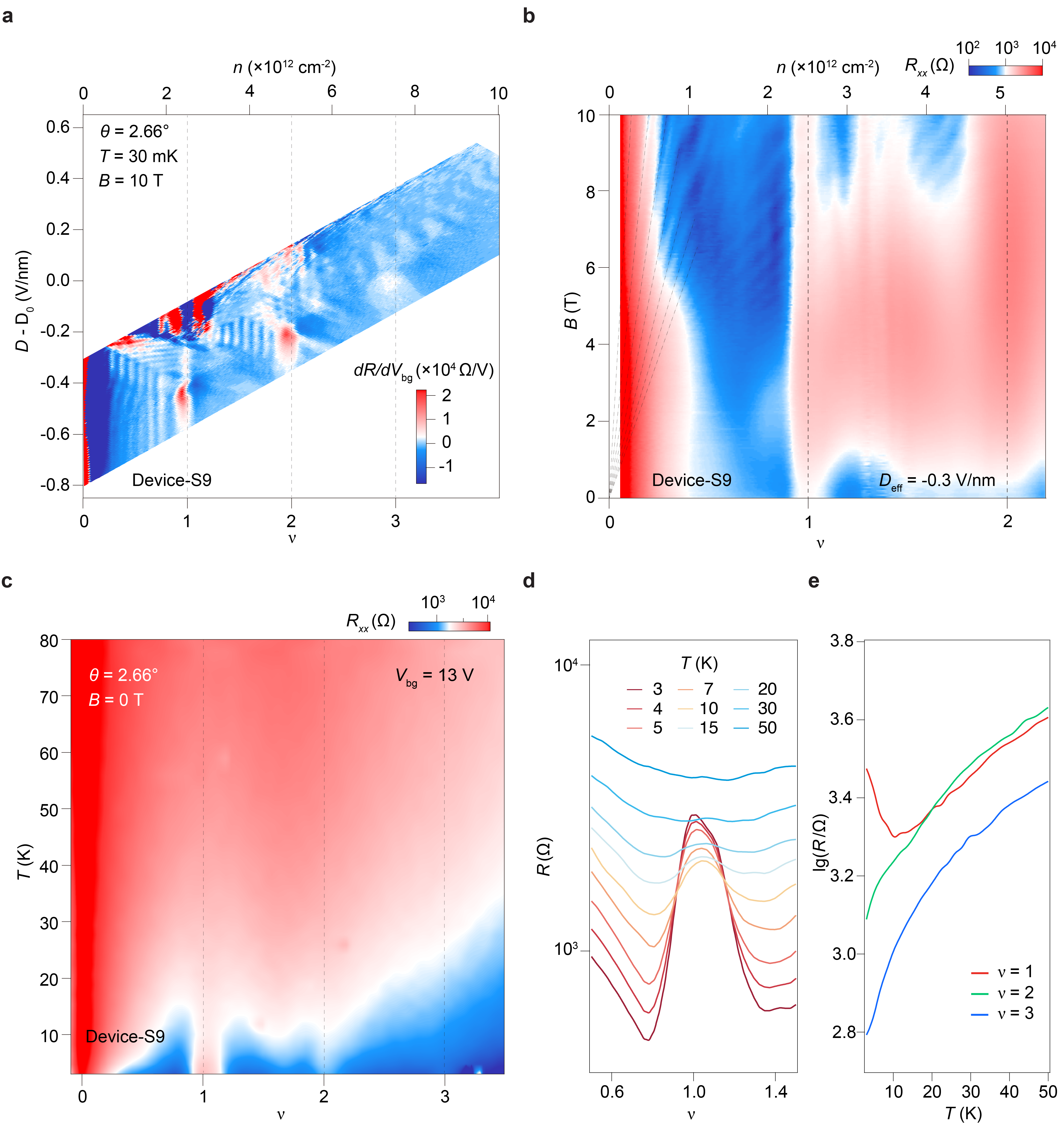}
  \caption{\textbf{Landau-level reconstruction and thermal evolution of correlated states in a $2.66^\circ$-twisted bilayer MoS$_2$ device.} \textbf{a}, Density derivative of the longitudinal resistance, $dR_{xx}/dV_{\rm bg}$, mapped in the displacement-field--filling-factor plane at $T=30$ mK and $B=10$ T. Taking the density derivative enhances weak density-dependent structures and reveals a sequence of Landau-level-like features in the charge-transfer region. \textbf{b}, Magnetic-field--filling-factor map of $R_{xx}$ measured at $D_{\rm eff}=-0.3$ V$\cdot$nm$^{-1}$.  \textbf{c}, Temperature-filling-factor map of $R_{xx}$ at zero magnetic field and fixed back-gate voltage $V_{\rm bg}=13$ V. The correlated resistance anomalies near integer moiré fillings are progressively suppressed upon warming. \textbf{d}, Density line cuts through the $\nu=1$ correlated resistance peak at several temperatures. The peak is strongest at low temperature and weakens continuously with increasing temperature. \textbf{e}, Temperature dependence of the correlated peaks at $\nu=1$, $2$ and $3$.}
  \label{fig:extended-data-4}
\end{figure*}
\clearpage

\end{document}